\documentclass{fizik}
\usepackage{amsmath}
\usepackage{amssymb}
\usepackage{psfrag}
\usepackage{graphicx}
\usepackage{psfig}
\usepackage{shortcuts}
\usepackage{latexsym}
\usepackage{longtable}
\usepackage[squaren,thickqspace,thickspace,cdot]{SIunits}

\vol{XX}
\pyear{2002}
\received{\today}
\author[TUNCER \& TUNCER]{
       {\bf En{\.{\i}}s Tuncer} 
        \thanks{{email: {\tt enis.tuncer@physics.org}};
         Currently at Department of Physics, University of Potsdam, Potsdam 14469 Germany.}\\
	{\it Chalmers University of Technology, 
         SE-412\ 96 Gothenburg Sweden}\\ 
	{\bf Emre Tuncer}\\
        {\it Monterey Design, 894 Ross Drive, Sunnyvale CA 94089 USA}}
\title{On complex permittivity of dilute random binary dielectric mixtures in two-dimensions
}
\begin{document}
\maketitle
\begin{abstract}
  Influence of number of particles considered in numerical simulations on complex dielectric permittivity of binary dilute dielectric mixtures in two-dimensions are reported. In the simulations, dodecagons (polygons with $12$-sides) were used to mimic disk-shaped inclusions. Using such an approach we were able to consider $16^2$ particles in a unit-square. The effective dielectric permittivity of the mixtures were calculated using the finite element at two different frequecies which were much higher and lower than the characteristic relaxation rate of the Maxwell-Wagner-Sillars polarization. The results were compared to an analytical solution.  It was found that the permittivity values at low frequencies were inside Wiener bounds however they violated the Hashin-Shtrikman bounds.
\end{abstract}

The complex permittivity of materials needs to be estimated in many applications~\cite{SihvolaBook,TuncerLic,TuncerPhD,Emre1992,Liu}. By using classical methods, {\em i.e.}, analytical techniques, effective medium approximation, {\em etc.}, the determination of the complex permittivity of dielectric materials is usually carried out for simple structures~\cite{SihvolaBook}. Moreover, the computed values should be verified by experiments or by means of other computational methods. Recently, the finite element ({\sc fe}) computer simulations~\cite{TuncerAcc1} has shown that disk-shaped inclusions in binary mixtures can be imitated by $n$-gons (polygon with $n$-sides) such that when $n\ge10$ the obtained electrical properties of the mixture were within $0.1\ \%$ error of the analytical solutions when the mixture was a dilute solution. In this paper, we investigate the electrical properties of a dilute binary mixture by considering the number of particles in the computational domain and compare the results with those obtained analytically~\cite{Emets98b}. 


In order to obtain random structures, an algorithm based on acception and rejection method (also known as random sequential adsorption) was constructed~\cite{Gervois}. In this approach, hard-disks with radii, $r$, were generated with random center coordinates in a unit square. Overlapping of disks both with outer boundaries of the unit square and with other disks were not allowed. The coordinate-values $x$ and $y$ were picked from a uniform random number generator where $(r,r)<(x,y)<(1-r,1-r)$. The first hard-disk was fixed, then, the number of hard-disks was incrimented. In each step, the condition for overlapping of the new generated disk with the already fixed disk(s) was checked. In case of overlapping, the disk was rejected, and a new one was picked from a new series of coordinates. 
In the process, obviously both the concentration of the hard-disks in the background medium and the number of disks should be kept constant. As mentioned previously, instead of using exact disk shapes and consuming lot of computational power, dodecagon 
(polygon with 12 sides) were used in the simulations.  The considered dodecagons were monodispersed.
The boundary conditions in the {\sc fe} were such that  reflections of the computational domain in the 4 facing-directions, von Neuman neighbors~\cite{Cellular} were used which was also Rayleigh's approach~\cite{Rayleigh}. The {\sc fe} simulation procedure has been presented previously elsewhere~\cite{sar97,Tuncer2001a,TuncerAcc1}.

In the simulations the concentration of the inclusion phase $q$ and number of inclusions in the computional domain were considered as variables. Only four concentration levels were included, $q=\{0.1,\ 0.2,\ 0.3,\ 0.4\}$. Starting with one inclusion, we have increased the number of inclusions $n$, in the computations.  The meshing of the region was arranged by the sizes of the dodecagons such that smallest meshing element was $1/4$  of the length of the  dodecagon side. This kind of procedure has allowed us to have enough number of discritization points in the inclusions. Computations with larger number of inclusions has increased the number of mesh elements, therefore, we were not able to obtain solutions for $n>16^2$ and $q\ge0.2$. For the $q=0.1$ case, the number of mesh elements were high enough when $n>12^2$. 

\begin{figure}[tp]
  \centering{
  \begin{tabular}{ccc}
    \includegraphics[width=2.1in]{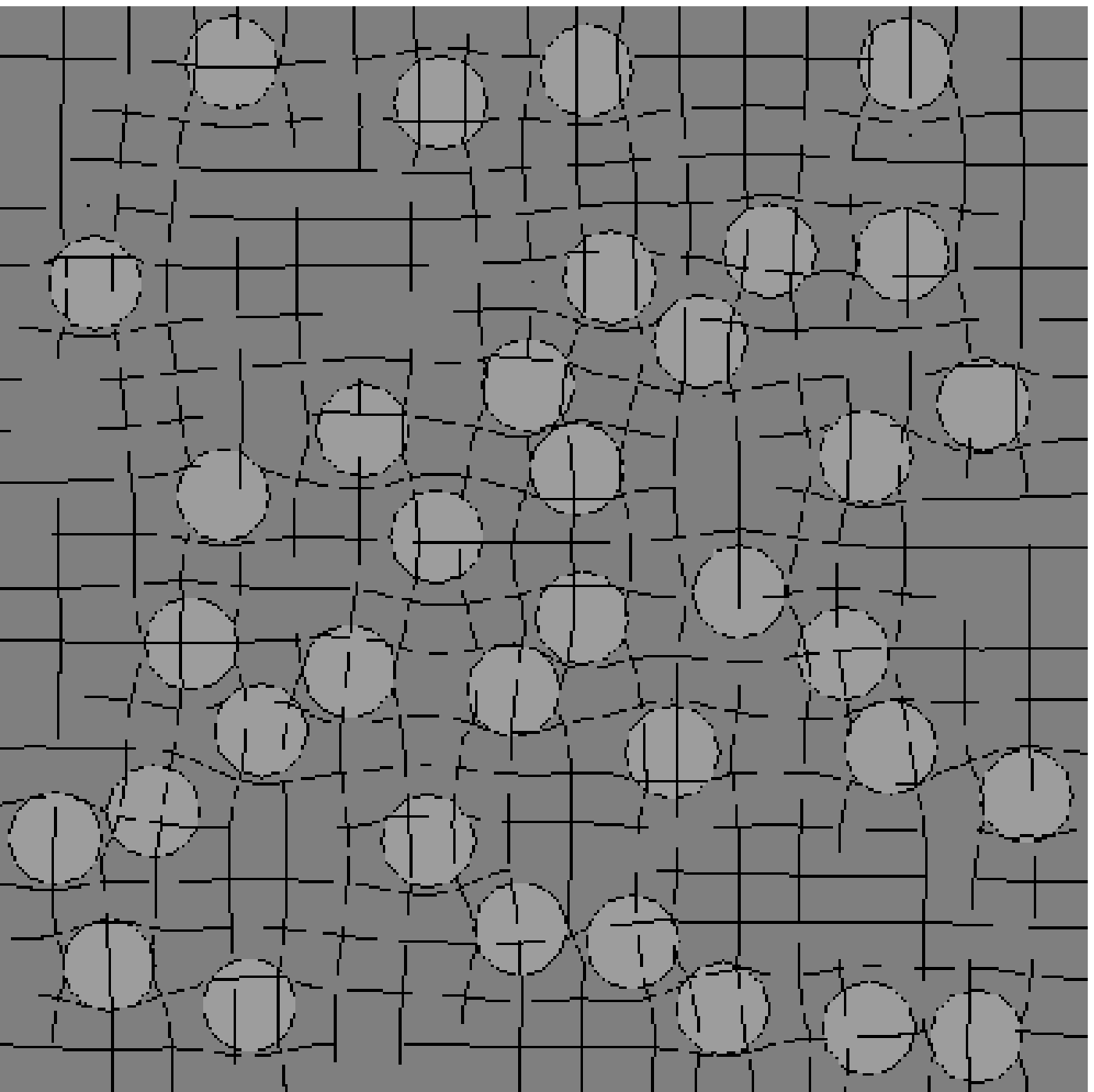}
      &\hspace{.3cm}
       &\includegraphics[width=2.1in]{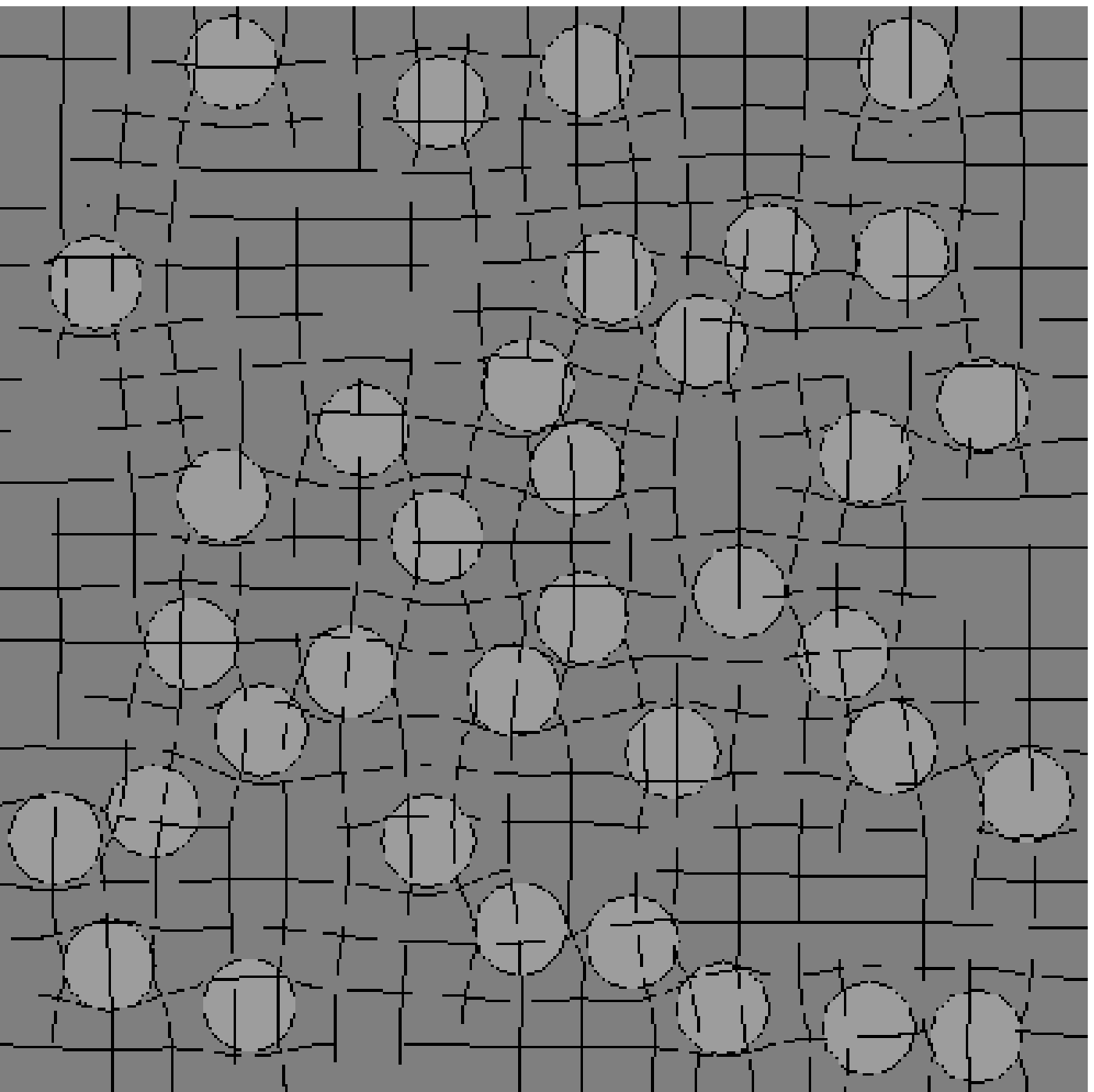}\\
       (a)& &(b)
  \end{tabular}}
  \caption{\label{fig:high1}
Potential (horizantal lines) and electric field (vertical lines) distributions at (a) lower and (b) higher frequencies than the characteristic interfacial polarization is observed. Number of inclusions are 36 and the concentration $q$ is 0.2.}
\end{figure}

 The electrical properties of the constituent media without any intrinsic polarizations were choosen such that the relaxation rate of the interfacial polarization in the composite medium~\cite{Maxwell1954,Wagner1914,Sillars1937} was around $1\ \hertz$. The complex dielectric permittivity of a medium as a function of angular frequency, $\omega=2\pi\nu$, can be expressed as,  
\begin{eqnarray}
  \label{eq:para}
  \widetilde\varepsilon_i(\omega)=\varepsilon_i+\frac{\sigma_i}{\imath\epsilon_0\omega}
\end{eqnarray}
where $\varepsilon_i$ is the dielectric permittivity of the medium at optical frequencies and $\sigma_i$ is the conductivity of the constituent. $\epsilon_0$ is the permittivity of the free space, $\epsilon_0=1/36\pi\ \nano\farad\per\meter$ and $\imath=\sqrt{-1}$. The permittivity and conductivity of the matrix phase, $\varepsilon_1$ and $\sigma_1$, were taken $2$ and $1\ \pico\siemens\per\meter$, respectively. The permittivity and conductivity of the dodecagon inclusions, $\varepsilon_2$ and $\sigma_2$, were $10$ and $100\ \pico\siemens\per\meter$, respectively. We have estimated the electrical properties $\varepsilon_\mega$, $\varepsilon_\micro$ and $\sigma$ where subscripts $\mega$ and $\micro$ represents the frequency. 

In Fig.~\ref{fig:high1}, the electric potential and electric field distributions of one of the considered case with $n=6^2$ and $q=0.2$ are presented. The difference between the two pictures are the one on the left is at low frequencies ($\nu=1\ \micro\hertz$) where the field distribution in the material is determined by the resistive properties of the media. Similarly, the picture on the right is at $\nu=1\ \mega\hertz$ where the capacitive properties of the media set the effective properties of the mixture.

\begin{figure}[tp]
  \centering{\includegraphics[width=5.5in]{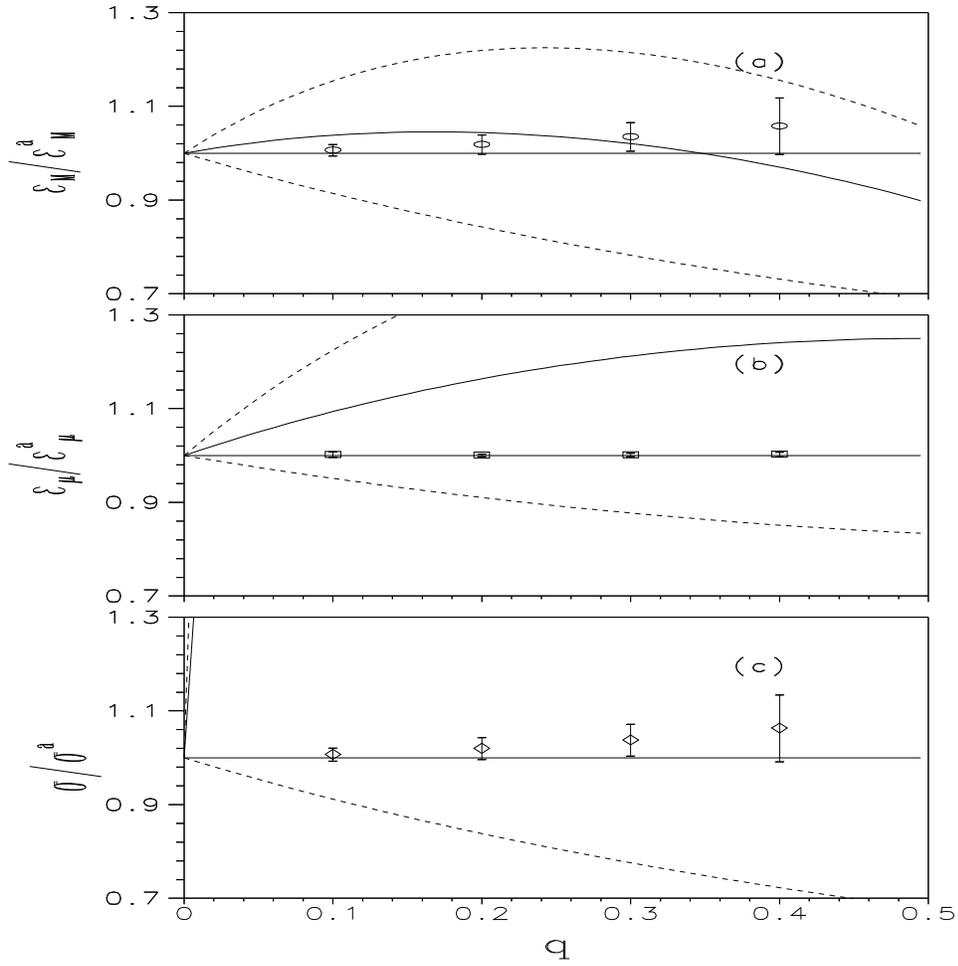}}
\caption{  \label{fig:q.vs.ratio}Ratio of electrical properties calculated and estimates of the analytical formula as a function of inclusion concentration $q$; (a) $\varepsilon_{\mega}$, (b) $\varepsilon_{\micro}$ and (c) $\sigma$. The dashed (\dashed) and solid (\full) lines represent the bounds of Wiener and Hashin-Shtrikman.}
\end{figure}
\begin{figure}[tp]
  \includegraphics[angle=-90,width=5.6in]{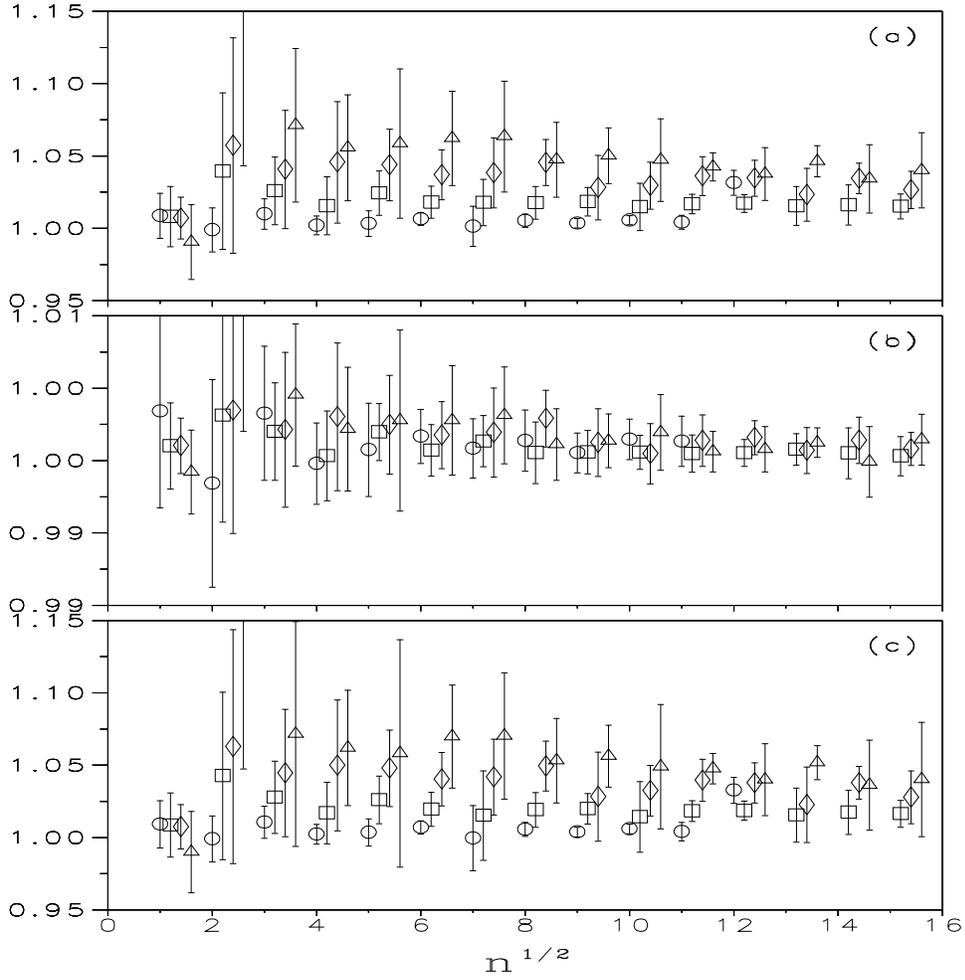}
\caption{\label{fig:dat1}Ratio of electrical properties calculated and estimates of the analytical formula as a function of square root number of inclusions $n$; (a) $\varepsilon_{\mega}$, (b) $\varepsilon_{\micro}$ and (c) $\sigma$. Different symbols represent the considered inclusion concentrations $q$; ($\bigcirc$) 0.1, ($\square$) 0.2, ($\Diamond$) 0.3 and ($\vartriangle$) 0.4.}
\end{figure}
The Maxwell-Garnett dielectric mixture formula~\cite{Maxwell_Garnett} is valid for two-dimensional dilute mixtures when the shape of the inclusions are disks~\cite{TuncerAcc1}. The effective complex dielectric permittivity of the mixture, $\widetilde\varepsilon_{\sf e}(\omega)$ can be found from 
\begin{equation}
  \label{eq:wiener_pier}
  \frac{\widetilde\varepsilon_{\sf e}(\omega)-\widetilde\varepsilon_1(\omega)}{\widetilde\varepsilon_{\sf e}(\omega)+(n-1)\widetilde\varepsilon_1(\omega)}=q\frac{\widetilde\varepsilon_2(\omega)-\widetilde\varepsilon_1(\omega)}{\widetilde\varepsilon_2(\omega)+(n-1)\widetilde\varepsilon_1(\omega)}
\end{equation}
where $n=2$ in two-dimensions\cite{Tuncer2001a}. In Fig.~\ref{fig:q.vs.ratio}, calculated electrical properties are presented as normalized to the values obtained from Eq.~(\ref{eq:wiener_pier}). In the figure, the bounds for dielectric permittivity and conductivity obtained from Wiener~\cite{wiener} and Hashin-Shtrikman~\cite{hashin} are also displayed as dashed (\dashed) and solid (\full) lines respectively. The Wiener bounds are obtained when $n=0$ and $n=\infty$ in Eq.~(\ref{eq:wiener_pier}). In two-dimensions, the Hashin-Shtrikman bounds are expressed as
\begin{equation}
  \label{eq:hashin}
  \widetilde\varepsilon_1+\frac{q}{(\widetilde\varepsilon_2-\widetilde\varepsilon_1)^{-1}+(1-q)(2\widetilde\varepsilon_1)^{-1}} \le \widetilde\varepsilon\le \widetilde\varepsilon_2+\frac{1-q}{(\widetilde\varepsilon_1-\widetilde\varepsilon_2)^{-1}+q(2\widetilde\varepsilon_2)^{-1}}
\end{equation}
The conductivity of the mixture was calculated from the dielectric losses, $\sigma=\Im[\widetilde\varepsilon''(\omega)\epsilon_0\omega]$ at $\omega=2\pi\ \micro\hertz$. It is significant from Fig.~\ref{fig:q.vs.ratio} that the lower bound of Hashin-Shtrikman is for disks since the obtained values ratios of the bounds and Eq.~(\ref{eq:wiener_pier}) with $n=2$ are equal. This proves that the considered dilute binary mixture was isotropic. The error bars in the figure or in other words the spread of the electrical properties indicate the importance of local field effects and interaction between inclusion particles such that at high concentrations the interaction between particles started to influence the electrical properties. 

When $\varepsilon$ at low frequencies are taken into consideration (Fig.~\ref{fig:q.vs.ratio}a), the upper limits of the two bounds have shown extraordinary behavior for higher concentrations of inclusions, the validity of Eq.~(\ref{eq:hashin}) was very narrow, $0<q<0.2$. The Wiener bounds are on the other hand valid for our simulations, however, for $q>\pi/4$ the Wiener bounds also show  extraordinary behavior. It is remarkable that $q=\pi/4$ is the limiting concentration for regular arrangement of inclusions on a square net.

The permittivity values at high frequencies are displayed in Fig.~\ref{fig:q.vs.ratio}b, and they were similar to the analytical formula in Eq.~(\ref{eq:wiener_pier}), and the spread of the data was very small compared to the conductivity values presented in Fig.~\ref{fig:q.vs.ratio}c. This was due to the ratio between the permittivity values considered, $\Re[\widetilde\varepsilon_1]/\Re[\widetilde\varepsilon_2]<\Im[\widetilde\varepsilon_1]/\Im[\widetilde\varepsilon_2]$~\cite{WebmanCohen}. For high frequency permittivity and ohmic conductivity both bounds were valid.

When the influence of the number of inclusions in the computational domain are considered, it was observed that the spread of the electrical values decreased as the number of inclusions increased. This behavior is displayed in Fig.~\ref{fig:dat1} again as the ratio between the computed values and Eq.~(\ref{eq:wiener_pier}). It was clear that the error-bars decreased by increasing the number of inclusions. There was a jump between the electrical properties when one-inclusion and four-inclusions systems were compared to each other. This was due to the change in the field distributions around inclusions which has pointed out the importance of boundary conditions in the simulations. At small scale systems with less number of inclusions, the boundary conditions are crucial since each particle is not affected by the local field distributions created by the neighboring inclusions. In addition, by increasing the number of inclusions in the computation domain, the probability of having larger number of neighbors in a radius is also increased when compared with a regular arrangement of inclusions~\cite{Tuncer2001a}.  This increase in the number of neighbors should in principle enhance the polarization of individual inclusions which is more realistic to real materials. This approach was desired but not selected due to difficulties arising in the field calculations, mainly the meshing procedure in the {\sc fe} method. The distances to nearest neighbors were not influenced by the number of inclusions if scaled with the inclusion radius. One could obviously use as many hard-disks as possible, after all it would be the most realistic condition. This could be done as proposed by considering a dodecagon instead of a circle. 
Finally, similar approach sould be applied to three dimensional problems to increase the number of considered particles in simulations.




\end{document}